\documentclass{IEEEtran4PSCC}
\ifCLASSINFOpdf
   \usepackage[pdftex]{graphicx}
\else
   \usepackage[dvips]{graphicx}
\fi
\usepackage[cmex10]{amsmath}
\usepackage[official]{eurosym}
\usepackage{threeparttable}
\usepackage{tabularx}
\usepackage{balance}

\usepackage{url}

\makeatletter
\let\old@ps@headings\ps@headings
\let\old@ps@IEEEtitlepagestyle\ps@IEEEtitlepagestyle
\def\psccfooter#1{%
    \def\ps@headings{%
        \old@ps@headings%
        \def\@oddfoot{\strut\hfill#1\hfill\strut}%
        \def\@evenfoot{\strut\hfill#1\hfill\strut}%
    }%
    \def\ps@IEEEtitlepagestyle{%
        \old@ps@IEEEtitlepagestyle%
        \def\@oddfoot{\strut\hfill#1\hfill\strut}%
        \def\@evenfoot{\strut\hfill#1\hfill\strut}%
    }%
    \ps@headings%
}
\makeatother

\psccfooter{%
        \parbox{\textwidth}{\hrulefill \\ \small{Accepted at 21st Power Systems Computation Conference} \hfill \small{Porto, Portugal --- June 29 -- July 3, 2020}}%
}

\begin{document}
%
\title{The Near-Optimal Feasible Space \\of a Renewable Power System Model}

\author{
  \IEEEauthorblockN{Fabian Neumann\\ Tom Brown}
  \IEEEauthorblockA{Institute for Automation and Applied Informatics \\
    Karlsruhe Institute of Technology (KIT)\\
    Karlsruhe, Germany\\
    \{fabian.neumann, tom.brown\}@kit.edu}
}


\maketitle

\begin{abstract}
  Models for long-term investment planning of the power system typically return
  a single optimal solution per set of cost assumptions. However, typically there
  are many near-optimal alternatives that stand out due to other attractive properties
  like social acceptance. Understanding features that persist across many cost-efficient
  alternatives enhances policy advice and acknowledges structural model uncertainties.
  We apply the modeling-to-generate-alternatives (MGA) methodology to systematically
  explore the near-optimal feasible space of a completely renewable European electricity
  system model. While accounting for complex spatio-temporal patterns, we allow simultaneous
  capacity expansion of generation, storage and transmission infrastructure subject to linearized
  multi-period optimal power flow. Many similarly costly, but technologically diverse solutions
  exist. Already a cost deviation of 0.5\% offers a large range of possible investments.
  However, either offshore or onshore wind energy along with some hydrogen storage and
  transmission network reinforcement appear essential to keep costs within 10\% of the optimum.
\end{abstract}


\thanksto{\noindent F.N. and T.B. gratefully acknowledge funding from the
  Helmholtz Association under grant no. VH-NG-1352. The responsibility for the contents lies with the authors.}

\section{Introduction}

As governments across the world are planning to increase the share of renewables,
energy system modeling has become a pivotal instrument for finding
cost-efficient future energy system layouts.
Energy system models formulate a cost minimization problem and typically return
a single optimal solution per set of input parameters (e.g. cost assumptions).

However, feasible but sub-optimal solutions may be preferable for reasons
that are not captured by model formulations because they are difficult to quantify
\cite{DeCarolis2016}.
Public acceptance of large infrastructure projects,
such as many onshore wind turbines or transmission network expansion,
ease of implementation,
land-use conflicts,
and regional inequality in terms of power supply
are prime examples of considerations which are
exogenous to most energy system models.
Bypassing such issues to enable a swift decarbonization of the energy system
may justify a limited cost increase.

Thus, providing just a singular optimal solution per scenario underplays
the degree of freedom in designing cost-efficient future energy systems.
Instead, presenting multiple alternative solutions and
pointing out features that persist across many near-optimal solutions
can remedy the lack of certainty in energy system models \cite{Pfenninger2014,DeCarolis2017}.
Communicating model results as a set of alternatives helps to identify
\textit{must-haves} (investment decisions common to all near-optimal solutions)
and \textit{must-avoids} (investment decisions not part of any near-optimal solution)
\cite{Hennen2017}.
The resulting boundary conditions can then inform political debate and support consensus building.

A common technique for determining multiple near-optimal solutions is called
Modeling to Generate Alternatives (MGA)
which uses the optimal solution as an anchor point to explore the surrounding
decision space for maximally different solutions \cite{DeCarolis2016}.
Other methods,
such as scenario analysis, global sensitivity analysis,
Monte Carlo analysis and stochastic programing,
that likewise address uncertainty in energy system modeling, concern \textit{parametric} uncertainty,
i.e. how investment choices change as cost assumptions are varied
\cite{DeCarolis2017,yue_review_2018,moret2017,shirizadeh2019}.
Conversely, MGA explores investment flexibility for a single set of
input parameters, by which it accounts for \textit{structural} uncertainty
and simplifications of model equations.
In consequence, MGA is a complement rather than a substitute
for methods sweeping across the parameter space.

\begin{table*}[]
  \caption{Literature Review: Studies Applying MGA to Energy System Models}
  \label{tab:literature}
  \begin{tabular}{|l|rrrrrrrrrr|}
    \hline
                                                   & Main          &         &       &                 &         & Max. GHG  & MGA       & Cost            & Near-optimal &      \\
    Source                                         & Sector        & Region  & Nodes & Snapshots       & Pathway & Reduction & Objective & Deviation       & Solutions    & LOPF \\ \hline
    Price et al. \cite{Price2017}                  & coupled (IAM) & global  & 16    & \textgreater{}1 & yes     & 50\%      & energy    & \textless{}10\% & 30           & no   \\
    DeCarolis et al. \cite{DeCarolis2011}          & electricity   & US      & 1     & 1               & no      & 85\%      & capacity  & \textless{}25\% & 9            & no   \\
    DeCarolis et al.  \cite{DeCarolis2016}         & electricity   & US      & 1     & 1               & yes     & 80\%      & energy    & \textless{}10\% & 28           & no   \\
    Li et al. \cite{Li2017}                        & electricity   & UK      & 1     & \textgreater{}1 & yes     & 80\%      & any       & \textless{}15\% & 800          & no   \\
    Sasse et al.  \cite{sasse_distributional_2019} & electricity   & CH      & 2,258 & 1               & no      & none      & energy    & \textless{}20\% & 2,000        & no   \\
    Trutnevyte et al. \cite{Trutnevyte2016}        & electricity   & UK      & 1     & 3               & no      & none      & any       & \textless{}23\% & 250,500      & no   \\
    Berntsen et al. \cite{berntsen_ensuring_2017}  & electricity   & CH      & 1     & 386             & no      & none      & any       & N/A             & 520          & no   \\
    Nacken et al.  \cite{nacken_integrated_2019}   & coupled       & DE      & 1     & 8,760           & no      & 95\%      & capacity  & \textless{}10\% & 1,025        & no   \\
    Hennen et al.   \cite{Hennen2017}              & urban energy  & generic & 1     & \textgreater{}1 & no      & none      & capacity  & \textless{}10\% & 384          & no   \\ \hline
    This study                                     & electricity   & Europe  & 100   & 4,380           & no      & 100\%     & capacity  & \textless{}10\% & 1,968        & yes  \\ \hline
  \end{tabular}
  \\~\\
  \textit{IAM--Integrated Assessment Model, GHG--greenhouse-gas, MGA--Modeling to Generate Alternatives, LOPF--Linear Optimal Power Flow,}\\
  \textit{UK--United Kingdom, CH--Switzerland, US--United States of America, DE--Germany}
  \vspace{-0.3cm}
\end{table*}

Evidence from previous work suggests many technologically diverse
solutions exist that result in similar total system costs for a sustainable
European power system \cite{schlachtberger_benefits_2017, schlachtberger_cost_2018}.
These two studies research the sensitivity of cost input parameters or the relevance
of transmission network expansion for low-cost power system layouts considering 30 regions.

Previous studies that applied MGA to long-term energy system planning problems or retrospective
analyses are reviewed in Table \ref{tab:literature}.
This work is the first to apply MGA to a European pan-continental electricity system
model which includes an adequate number of regions and operating conditions to reflect
the complex spatio-temporal patterns shaping cost-efficient investment strategies
in a fully renewable system.
Furthermore, the co-optimization of generation, storage and transmission
infrastructure subject to linear optimal power flow (LOPF) constraints is unique
for MGA applications.

The goal of this work is to systematically explore the wide array of similarly
costly but diverse technology mixes for the European power system,
and derive a set of rules that must be satisfied to keep costs within pre-defined ranges.
Additionally, we investigate how the extent of investment flexibility changes
as we apply more ambitious greenhouse gas (GHG) emission reduction targets up to a
complete decarbonization and allow varying levels of relative cost increases.

The remainder of the paper is structured as follows:
Section \ref{sec:methodology} guides through the problem formulation,
the employed variant of MGA, sources of model input data, and the experimental setup.
The results are presented and discussed from different perspectives in Section \ref{sec:results}
and critically appraised in Section \ref{sec:appraisal}.
The work is concluded in Section \ref{sec:conclusion}.

\section{Methodology}
\label{sec:methodology}

\subsection{Problem formulation for long-term power system planning}

The objective of long-term power system planning is to minimize the total
annual system costs, comprising annualised\footnote{The annuity factor $\frac{1-(1+\tau)^{-n}}{\tau}
  $converts the overnight investment of an asset to annual payments considering its
  lifetime $n$ and cost of capital $\tau$.} capital costs $c_*$ for investments at locations $i$
in generator capacity $G_{i,r}$ of technology $r$, storage capacity $H_{i,s}$ of technology $s$, and transmission line capacities
$F_{\ell}$, as well as the variable operating costs $o_*$ for generator dispatch $g_{i,r,t}$:
\begin{align}
  \min_{G,H,F,g} \; f(G,H,F,g) \;=\; \min_{G,H,F,g} \quad \left[\sum_{i,r} c_{i,r}\cdot G_{i,r} +  \right. \nonumber \\
  \left. \sum_{i,s} c_{i,s}\cdot H_{i,s} + \sum_{\ell}c_{\ell}\cdot F_{\ell}+\sum_{i,r,t}w_t\cdot o_{i,r} \cdot g_{i,r,t} \right]
\end{align}
where representative snapshots $t$ are weighted by $w_t$ such that their total duration
adds up to one year; \mbox{$\sum_{t=1}^{T}w_t=365\cdot 24\text{h}=8760\text{h}$}.
The objective function is subject to a set of linear constraints,
including multi-period linear optimal power flow (LOPF) equations, resulting in a convex linear program (LP).

The capacities of generation, storage and transmission infrastructure are
constrained by their geographical potentials:
\begin{align}
  \underline{G}_{i,r}  &  & \leq &  & G_{i,r}  &  & \leq &  & \overline{G}_{i,r}  & \qquad\forall i, r \\
  \underline{H}_{i,s}  &  & \leq &  & H_{i,s}  &  & \leq &  & \overline{H}_{i,s}  & \qquad\forall i, s \\
  \underline{F}_{\ell} &  & \leq &  & F_{\ell} &  & \leq &  & \overline{F}_{\ell} & \qquad\forall \ell
\end{align}

The dispatch of a generator may not only be constrained by its rated capacity but also by the availability of
variable renewable energy, which may be derived from reanalysis weather data.
This can be expressed as a time- and location-dependent availability
factor $\overline{g}_{i,r,t}$, given in per-unit of the generator's capacity:
\begin{align}
  0 &  & \leq &  & g_{i,r,t} &  & \leq &  & \overline{g}_{i,r,t} G_{i,r} & \qquad\forall i, r, t
\end{align}

The dispatch of storage units is split into two positive variables;
one each for charging $h_{i,s,t}^+$ and discharging $h_{i,s,t}^-$.
Both are limited by the power rating $H_{i,s}$ of the storage units.
\begin{align}
  0 &  & \leq &  & h_{i,s,t}^+ &  & \leq &  & H_{i,s} & \qquad\forall i, s, t \\
  0 &  & \leq &  & h_{i,s,t}^- &  & \leq &  & H_{i,s} & \qquad\forall i, s, t
\end{align}
This formulation does not prevent simultaneous charging and discharging,
in order to maintain the computational benefit of a convex feasible space.
The energy levels $e_{i,s,t}$ of all storage units have to be consistent with the dispatch in all hours.
\begin{align}
  e_{i,s,t} =\: & \eta_{i,s,0}^{w_t} \cdot e_{i,s,t-1} + w_t \cdot h_{i,s,t}^\text{inflow} - w_t \cdot h_{i,s,t}^\text{spillage} & \quad\forall i, s, t \nonumber \\
                & + \eta_{i,s,+} \cdot w_t \cdot h_{i,s,t}^+ - \eta_{i,s,-}^{-1} \cdot w_t \cdot h_{i,s,t}^-
\end{align}
Storage units can have a standing loss $\eta_{i,s,0}$, a charging efficiency $\eta_{i,s,+}$, a discharging efficiency $\eta_{i,s,-}$,
natural inflow $h_{i,s,t}^\text{inflow}$ and spillage $h_{i,s,t}^\text{spillage}$.
The storage energy levels are assumed to be cyclic
\begin{align}
  e_{i,s,0} = e_{i,s,T} \qquad\forall i, s
\end{align}
and are constrained by their energy capacity
\begin{align}
  0 &  & \leq &  & e_{i,s,t} &  & \leq &  & \overline{T}_s \cdot H_{i,s} & \qquad\forall i, s, t.
\end{align}
To reduce the number of decisison variables, we tie the energy storage volume to
power ratings using a technology-specific parameter $\overline{T}_s$
that describes the maximum duration a storage unit can discharge at full power rating.

Kirchhoff's Current Law (KCL) requires local generators and storage units as well as
incoming or outgoing flows $f_{\ell,t}$ of incident transmission lines $\ell$
to balance the inelastic electricity demand $d_{i,t}$ at each location $i$ and snapshot $t$
\begin{align}
  \sum_r g_{i,r,t} + \sum_s h_{i,s,t} + \sum_\ell K_{i\ell} f_{\ell,t} = d_{i,t} \qquad\forall i,t
\end{align}
where $K_{i\ell}$ is the incidence matrix of the network.

Kichhoff's Voltage Law (KVL) imposes further constraints on the flow of passive AC lines.
Using linearized load flow assumptions, the voltage angle difference around every closed cycle in the
network must add up to zero. This constrained can be formulated using a cycle basis
of the network graph where the independent cycles $c$ that span the cycle space are expressed as
directed linear combinations of lines $\ell$ in a cycle incidence matrix $C_{\ell c}$ \cite{cycleflows}.
This leads to the constraint
\begin{align}
  \sum_\ell C_{\ell c} \cdot x_\ell \cdot f_{\ell,t} = 0 \qquad\forall c,t
\end{align}
where $x_\ell$ is the series inductive reactance of line $\ell$.
The controllable HVDC links are not affected by this constraint.

All line flows $f_{\ell,t}$ are also limited by their capacities $F_\ell$
\begin{align}
  |f_{\ell,t}| \leq \overline{f}_{\ell} F_{\ell} & \qquad\forall \ell, t,
\end{align}
where $\overline{f}_\ell$ acts as a per-unit security margin on the line capacity.

Finally, total CO$_2$ emissions may not exceed a target level $\Gamma_{\text{CO}_2}$.
The emissions are determined from the time-weighted generator dispatch $ w_t \cdot g_{i,r,t}$ using the specific emissions $\rho_r$ of fuel $r$
and the generator efficiencies $\eta_{i,r}$:
\begin{align}
  \sum_{i,r,t}  \rho_r \cdot \eta_{i,r}^{-1} \cdot w_t \cdot g_{i,r,t} \leq \Gamma_{\text{CO}_2}.
\end{align}

Note, that this formulation does not include pathway optimization (i.e.~no sequences of investments),
but searches for a cost-optimal layout corresponding to a given GHG emission reduction level.
For capacity expansion planning,
it assumes perfect foresight for the reference year based on which capacities are optimized.
Additional aspects such as reserve power, system stability, or robust scheduling have not been considered here.
The optimization problem is implemented in the open-source modeling framework PyPSA \cite{brown_pypsa_2018}.

\subsection{Modeling to generate alternatives (MGA)}

\begin{figure}
  \centering
  \includegraphics[width=0.95\columnwidth, trim=0cm 0.3cm 0cm 0.3cm, clip]{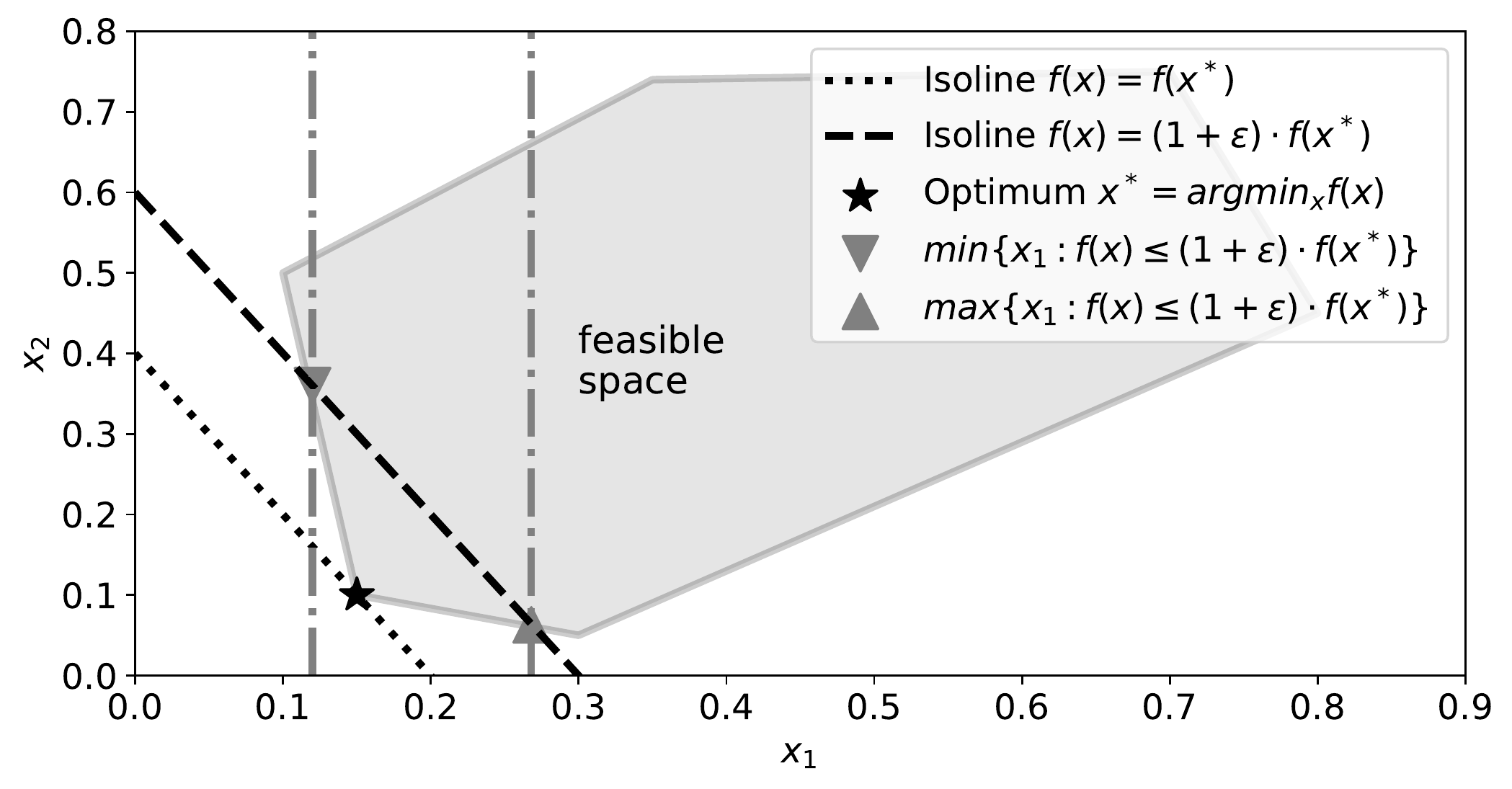}
  \caption{Illustration of the near-optimal feasible space and the modeling to generate alternatives (MGA) methodology for a two-dimensional problem for the search-directions relating to dimension $x_1$.}
  \label{fig:mga}
  \vspace{-0.3cm}
\end{figure}

As shown in Fig. \ref{fig:mga}, following a run of the original model, the objective function is encoded as a constraint
such that the original feasible space is limited by the optimal objective value $f^*$ plus some acceptable relative cost increase $\epsilon$.
\begin{equation}
  f(G,H,F,g)\leq (1+\epsilon) \cdot f^*
\end{equation}
Other than preceding studies, this paper pursues a more structured approach
to MGA to span the near-optimal feasible space.
The search directions are not determined by the Hop-Skip-Jump (HSJ) algorithm
that seeks to minimize the weighted sum of variables of previous solutions \cite{yue_review_2018},
but by pre-defined groups of investment variables.
Consequently, the new objective function becomes to variously minimize and maximize
sums of subsets of generation, storage and transmission capacity expansion
variables given the $\epsilon$-cost constraint.

The groups can be formed by region and by technology.
Examples for thought-provoking search directions are to minimize the sum of onshore wind
capacity in Germany or the total volume of transmission expansion (cf. Section \ref{sec:setup}).

This process yields boundaries within which all near-optimal solutions are contained
and can be interpreted as a set of rules that must be followed to become nearly cost-optimal.

In fact, by arguments of convexity, it can be shown that near-optimal
solutions exist for all values of a group's total capacity between
their corresponding minima and maxima.
The original problem is convex as it classifies as a linear program.
Neither adding the linear $\epsilon$-cost constraint nor introducing an
auxiliary variable $z$ that represents the sum of the group of variables alter this characteristic.
Hence, for any total $z\in[z_{\text{min}}, z_{\text{max}}]$ a near-optimal
solution exists, however not for any combination of its composites leading to this total.

\begin{table*}%
  \begin{center}

    \begin{threeparttable}
      \caption{Assumptions for Techno-Economic Input Parameters}
      \label{tab:costsassumptions}
      \begin{tabular}{|lrrrrrrrr|}
        \hline

        Technology\tnote{e}      & Investment         & Fixed O\&M   & Marginal      & Lifetime & Efficiency                 & Investment   & $\overline{T}$\tnote{f} & Source                                \\
                                 & [\euro/kW]         & [\euro/kW/a] & [\euro/MWh]   & [a]      & [-]                        & [\euro/kWh]  & [h]                     &                                       \\

        \hline

        Onshore Wind             & 1330               & 33           & 2.3           & 25       & 1                          &              &                         & DEA \cite{DEA}                        \\
        Offshore Wind (AC)       & 1890               & 44           & 2.7           & 25       & 1                          &              &                         & DEA \cite{DEA}                        \\
        Offshore Wind (DC)       & 2040               & 47           & 2.7           & 25       & 1                          &              &                         & DEA \cite{DEA}                        \\
        Solar                    & 600                & 25           & 0.01          & 25       & 1                          &              &                         & Schröder et al. \cite{schroeder2013}  \\
        Run of River             & 3000               & 60           & 0             & 80       & 0.9                        &              &                         & Schröder et al. \cite{schroeder2013}  \\\hline
        OCGT\tnote{a}            & 400                & 15           & 58.4\tnote{b} & 30       & 0.39                       &              &                         & Schröder et al. \cite{schroeder2013}  \\
        CCGT\tnote{a}            & 800                & 20           & 47.2\tnote{b} & 30       & 0.5                        &              &                         & Schröder et al. \cite{schroeder2013}  \\ \hline
        Hydrogen                 & 689                & 24           & 0             & 20       & 0.8 $\cdot$ 0.58\tnote{c}  & 8.4          & 168                     & Budischak et al. \cite{budischak2013} \\
        Battery                  & 310                & 9            & 0             & 20       & 0.81 $\cdot$ 0.81\tnote{c} & 144.6        & 6                       & Budischak et al. \cite{budischak2013} \\
        Pumped Hydro             & 2000               & 20           & 0             & 80       & 0.75                       & N/A\tnote{d} & 6                       & Schröder et al. \cite{schroeder2013}  \\
        Hydro Reservoir          & 2000               & 20           & 0             & 80       & 0.9                        & N/A\tnote{d} & fixed                   & Schröder et al. \cite{schroeder2013}  \\ \hline
        Transmission (submarine) & {2000 \euro /MWkm} & 2\%          & 0             & 40       & 1                          &              &                         & Hagspiel et al. \cite{Hagspiel}       \\
        Transmission (overhead)  & {400 \euro /MWkm}  & 2\%          & 0             & 40       & 1                          &              &                         & Hagspiel et al. \cite{Hagspiel}       \\
        \hline
      \end{tabular}

      \begin{tablenotes}
        \item [a] Gas turbines have a CO$_2$ emission intensity of 0.19 t/MW$_{\text{th}}$.
        \item [b] This includes fuel costs of 21.6 \euro/MWh$_{\text{th}}$.
        \item [c] The storage round-trip efficiency consists of charging and discharging efficiencies $\eta_+ \cdot \eta_-$.
        \item [d] The installed facilities are not expanded in this model and are considered to be amortized. %
        \item [e] For all technologies a discount rate of 4\% is assumed.
        \item [f] This relates a storage unit's energy capacity to its power capacity; it is the maximum duration the storage unit can discharge at full power capacity.

      \end{tablenotes}
    \end{threeparttable}
  \end{center}
  \vspace{-0.5cm}
\end{table*}

\subsection{Model input data}

The exploration of the near-optimal feasible space is executed for the open model
dataset PyPSA-Eur of the European power system with a spatial resolution of 100
nodes and a temporal resolution of 4380 snapshots (two-hourly for a full year) \cite{Horsch2018}.
The chosen levels of geographical and temporal aggregation reflect,
at the upper end, the computational limits to calculate a large ensemble of
near-optimal solutions and, at the lower end, the minimal requirements to expose
transmission bottlenecks and account for spatially and temporally varying
renewable potentials with passable detail \cite{Hoersch2017d, pfenninger2017}.

Following a greenfield approach (with the exception of the transmission
grid and hydropower installations), we allow simultaneous capacity
expansion of transmission lines, HVDC links and various types of storage
units and generators: solar photovoltaics, onshore wind turbines, offshore
wind turbines with AC or DC grid connections, battery storage, hydrogen
storage and, ultimately, open- and combined cycle gas turbines (OCGT/CCGT)
as sole fossil-fueled plants.

Run-of-river and pumped-hydro capacities are
not extendable due to assumed geographical constraints.
All other generators and storage units can be built at any location
up to their geographical potentials. The corridors for new
HVDC links, limited to 30 GW, are taken from the TYNDP 2018 \cite{tyndp2018}.
Individual AC transmission line capacities may be expanded continuously up
to four times their current capacity, but not reduced.
We further assume that the annual electricity demand for the power sector does
not deviate substantially from today's levels.
Given the densely meshed and spatially aggregated transmission system,
we do not add new expansion corridors but constrain options to reinforcement via parallel AC lines.
To approximate $N-1$ security, the effective transfer capacity of transmission
lines is restricted to 70\% of their nominal rating \cite{Horsch2018}.
The dependence of line capacity expansion on line impedance is addressed
in a sequential linear programing approach \cite{neumann_heuristics_2019}.
This relaxation is justified as it removes the excessive computational burden of integer programming,
while yielding equally accurate solutions given tolerated optimality gaps in discrete problems
and other more decisive model condensations (e.g. network clustering) \cite{neumann_heuristics_2019}.
Full details on the workflow of PyPSA-Eur and processing the underlying
datasets can be found in \cite{Horsch2018}.
Cost assumptions and further techno-economic input parameters are
listed in Table \ref{tab:costsassumptions}.

\begin{figure}
  \begin{center}
    \begin{scriptsize}
      (i) Optimal Transmission Network Expansion: $\epsilon=0\%$ / GHG -100\%\\
      \includegraphics[width=0.999\columnwidth, trim=0cm 6.4cm 0cm 1.4cm, clip]{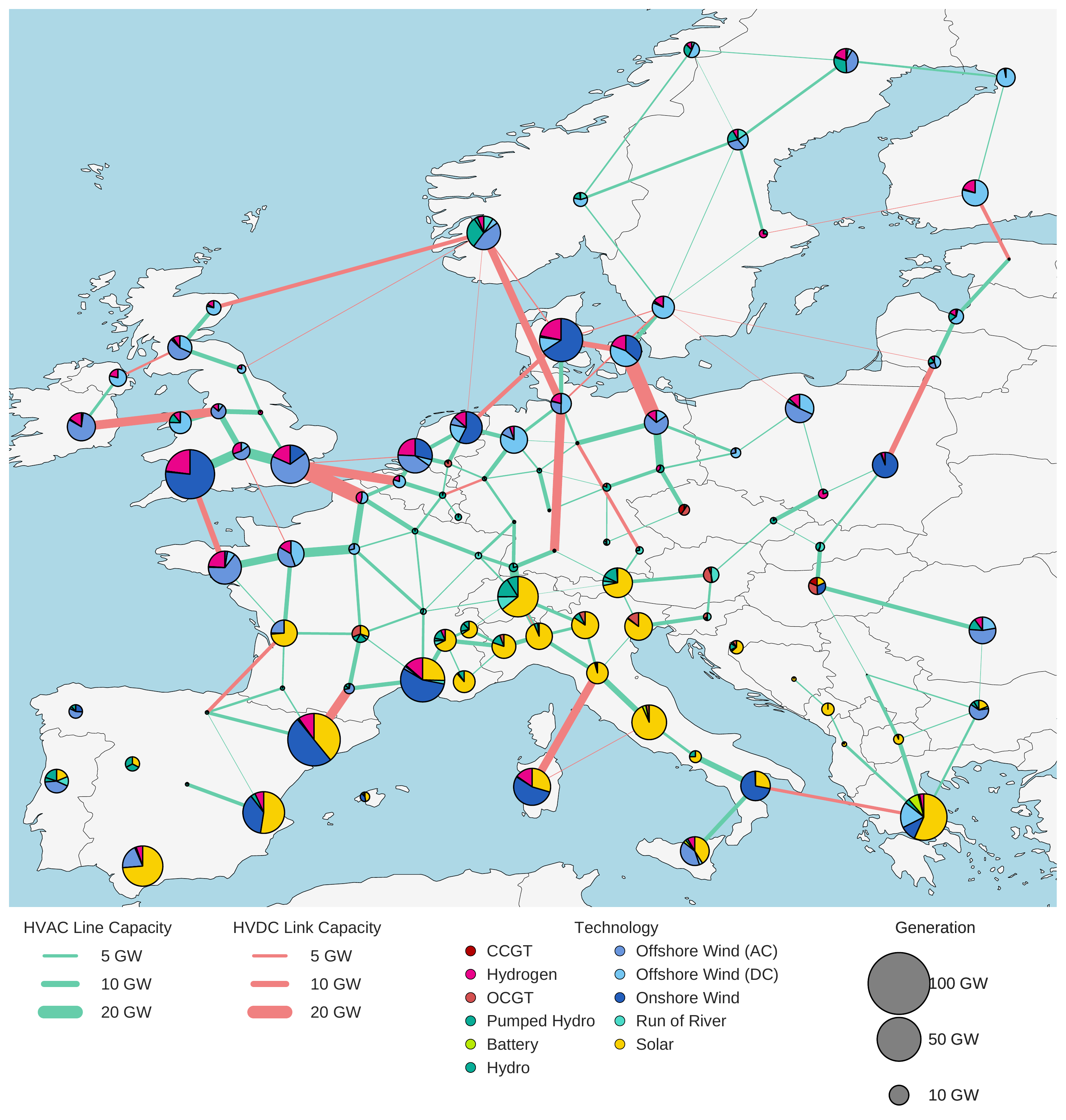}\\
      (ii) Minimize Transmission Network Expansion: $\epsilon=10\%$ / GHG -100\%\\
      \includegraphics[width=0.990\columnwidth, trim=0cm 1cm 0cm 1.4cm, clip]{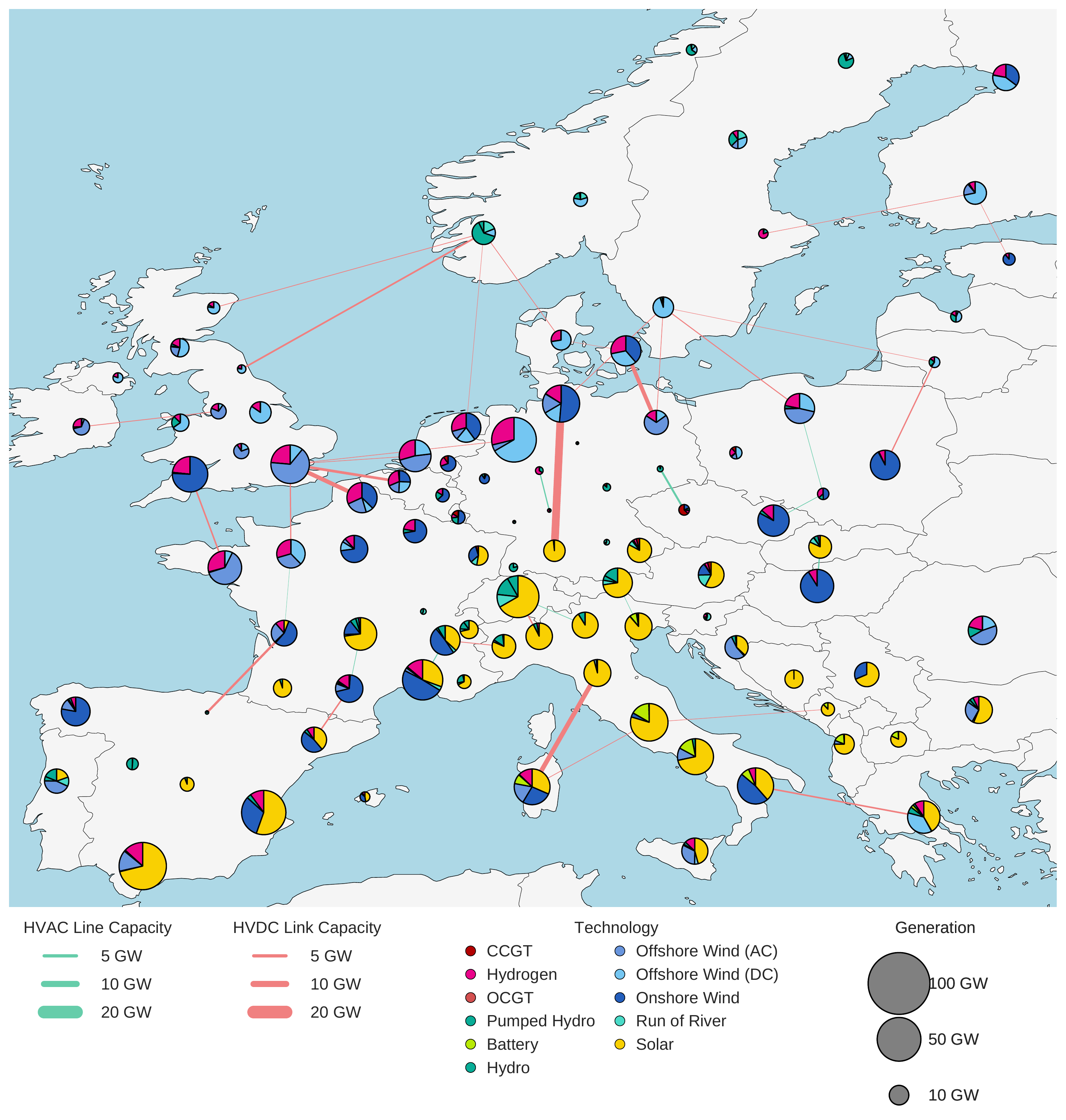}\\
    \end{scriptsize}
  \end{center}
  \vspace{-0.4cm}
  \caption{Maps of transmission line expansion and regional generator and storage capacities for a 100\% renewable power system for the (i) optimal solution and (ii) minimal transmission volume within a 10\% cost increase.
  }
  \label{fig:maps-tmin}
  \vspace{-0.3cm}
\end{figure}

\subsection{Experimental setup}
\label{sec:setup}

The MGA analysis is run within a parallelized workflow for different deviations
\mbox{$\epsilon \in \{\text{0.5\%, 1\%, 2\%, 3\%, 4\%, 5\%, 7.5\%, 10\%}\}$}
from the cost-optimal solution and for system-wide greenhouse-gas emission
reduction targets of 80\%, 95\% and 100\% compared to emission levels in 1990.
This allows to follow the propagation of investment flexibility for increasing
optimality tolerances and more ambitious climate protection plans.
The alternative objectives are to minimize and maximize the generation capacity of all
(i) wind turbines,
(ii) onshore wind turbines,
(iii) offshore wind turbines,
(iv) solar panels, and
(v) natural gas turbines.
Moreover, we search for the minimal and maximal deployment of
(vi) hydrogen storage,
(vii) battery storage, and
(viii) power transmission infrastructure.
This setup yields 384 near-optimal solutions.
On average, each problem required 6.5 hours and 31 GB of memory to solve with the Gurobi solver.

For slacks $\epsilon \in \{\text{1\%, 5\%, 10\%}\}$ and a 95\% emission reduction
target a 3-hourly resolved model is run for country-wise minima and maxima of the
investment groups above, resulting in additional 1584 near-optimal solutions.
On average, each problem required 3.5 hours and 22 GB of memory to solve.

\section{Results and Discussion}
\label{sec:results}

\begin{table}
  \centering
  \caption{Statistics on Optimal Solutions for Different GHG Emission Reduction Levels}
  \label{tab:results}
  \begin{tabular}{|l|r|r|r|}
    \hline
    \hfill \textbf{GHG Emissions}     & \textbf{-100\%} & \textbf{-95\%} & \textbf{-80\%} \\
    \textbf{Generation {[}TWh{]}}     &                 &                &                \\ \hline
    Onshore Wind                      & 750 (24\%)      & 421  (14\%)    & 423  (15\%)    \\
    Offshore Wind (AC)                & 886 (28\%)      & 568 (19\%)     & 297 (10\%)     \\
    Offshore Wind (DC)                & 873 (27\%)      & 1032 (34\%)    & 769 (27\%)     \\
    Solar                             & 502 (16\%)      & 605 (20\%)     & 381 (13\%)     \\
    Run of River                      & 150 (5\%)       & 153 (6\%)      & 154 (5\%)      \\
    CCGT (gas)                        & 0 (0\%)         & 171 (6\%)      & 761 (26\%)     \\
    OCGT (gas)                        & 0 (0\%)         & 41 (1\%)       & 105 (4\%)      \\ \hline
    \textbf{Transmission [TWkm]}      & 504 (+71\%)     & 458 (+55\%)    & 368 (+25\%)    \\
    \textbf{Load [TWh]}               & 3,138           & 3,138          & 3,138          \\
    \textbf{Total Cost [bn \euro/a]}  & 246             & 207            & 165            \\
    \textbf{Total Cost [\euro{}/MWh]} & 78.4            & 66.1           & 52.6           \\ \hline
  \end{tabular}
  \vspace{-0.3cm}
\end{table}

\subsection{Optimal solutions}

Before delving into near-optimal solutions, we first outline the characteristics
of the optimal solutions for different emission reduction levels (cf. Table \ref{tab:results}).
A system optimized for a 100\% emission reduction is strongly dominated by wind energy.
More than half of the electricity is supplied by offshore wind installations.
Onshore wind turbines provide another quarter.
In contrast, photovoltaics account for only 16\% of electricity generation.
Strikingly, a system targeting a 95\% reduction in greenhouse gases uses significantly
less onshore wind generators but more solar energy in comparison to a
completely decarbonized system, while  keeping the share of offshore wind generation constant.
Thus, for the last mile from 95\% to 100\% more onshore wind generation is
preferred to phase out the last remaining natural-gas-fired power plants.
The total system costs scale non-linearly with more tight emission caps.
Achieving an emission reduction of 95\% is roughly a quarter more expensive than a reduction
by 80\%, while a zero-emission system is almost 50\% more expensive.

Also, the map in Fig. \ref{fig:maps-tmin}-(i) shows the optimal regional distribution of the
capacities of power system components for a fully renewable European power system.
Generation hubs tend to form along the coasts of North, Baltic and Mediterranean Sea, whereas
inland regions produce little electricity. Expectedly, solar energy is the dominant carrier
in the South, while wind energy prevails close to the coasts of the North Sea and the Baltic Sea.
Most grid expansion can be found in Germany, France and the United Kingdom and
individual HVDC links are built with capacities of up to 30 GW.
The routes and capacities of HVDC links are well correlated with the placement of wind farms.

\subsection{The near-optimal feasible space}

\begin{figure}
  \centering
  \includegraphics[width=\columnwidth]{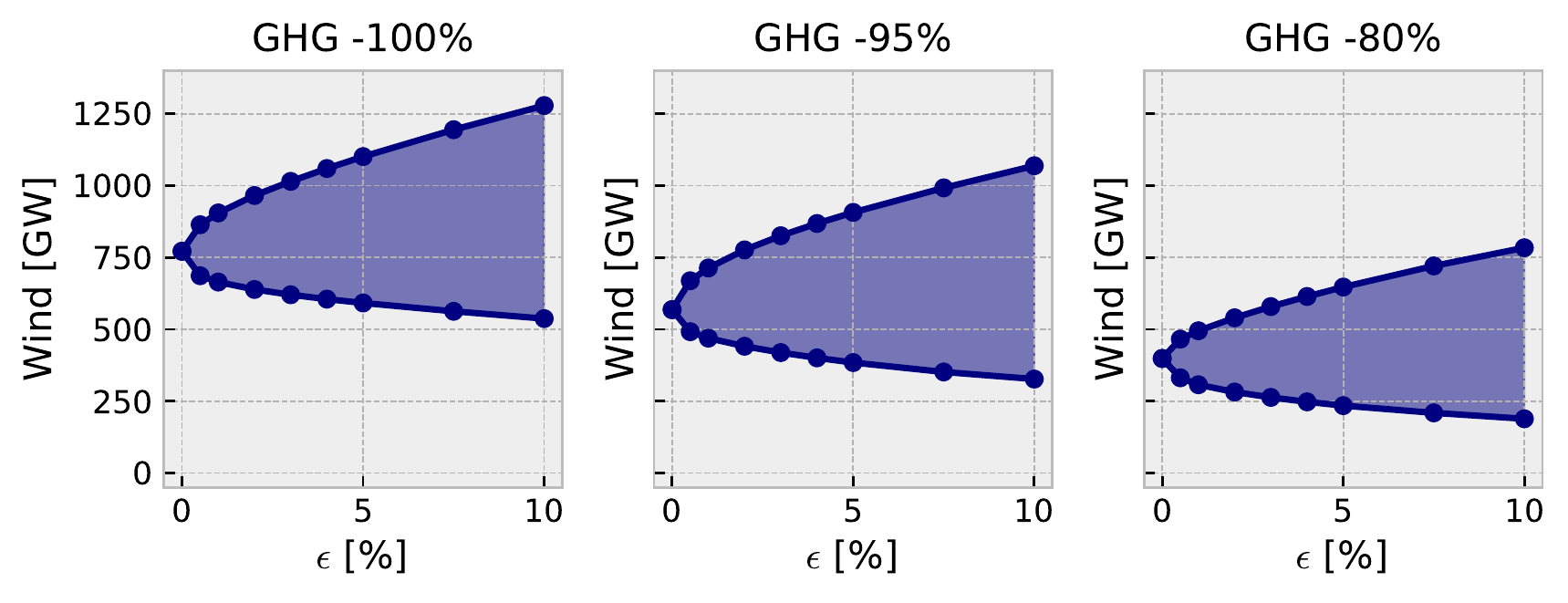}
  \includegraphics[width=\columnwidth]{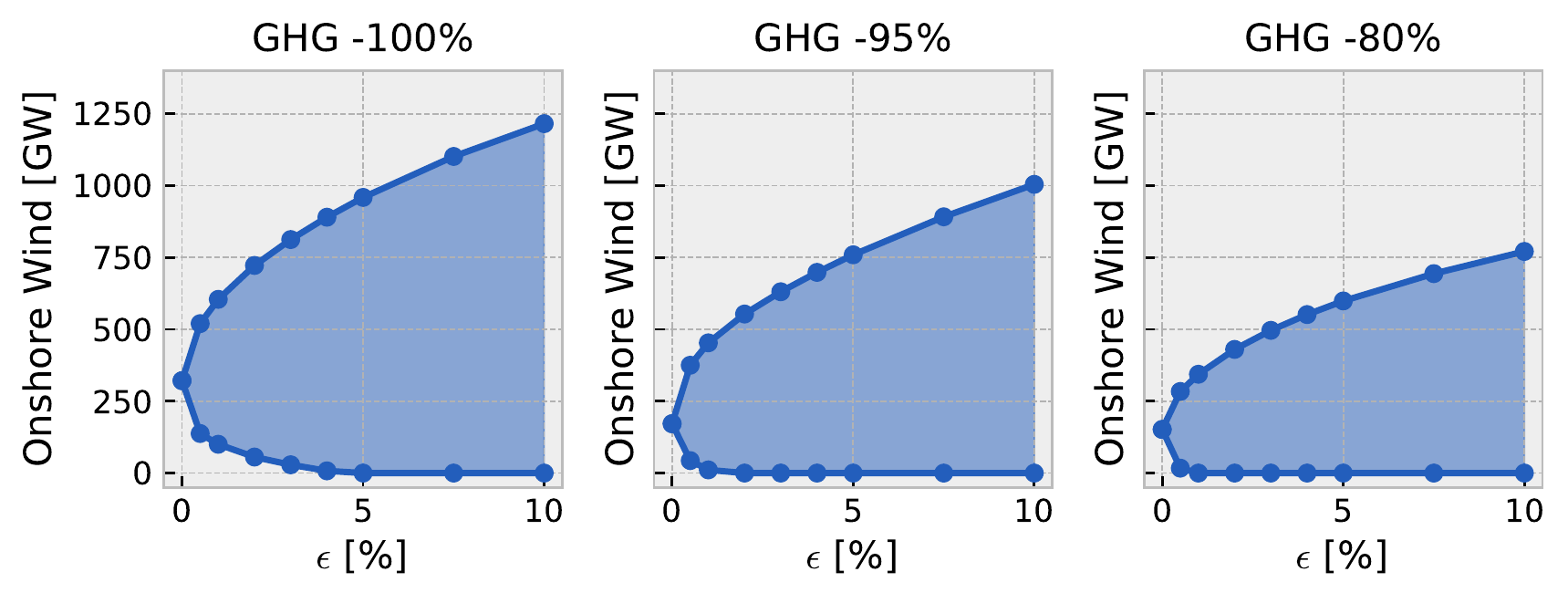}
  \includegraphics[width=\columnwidth]{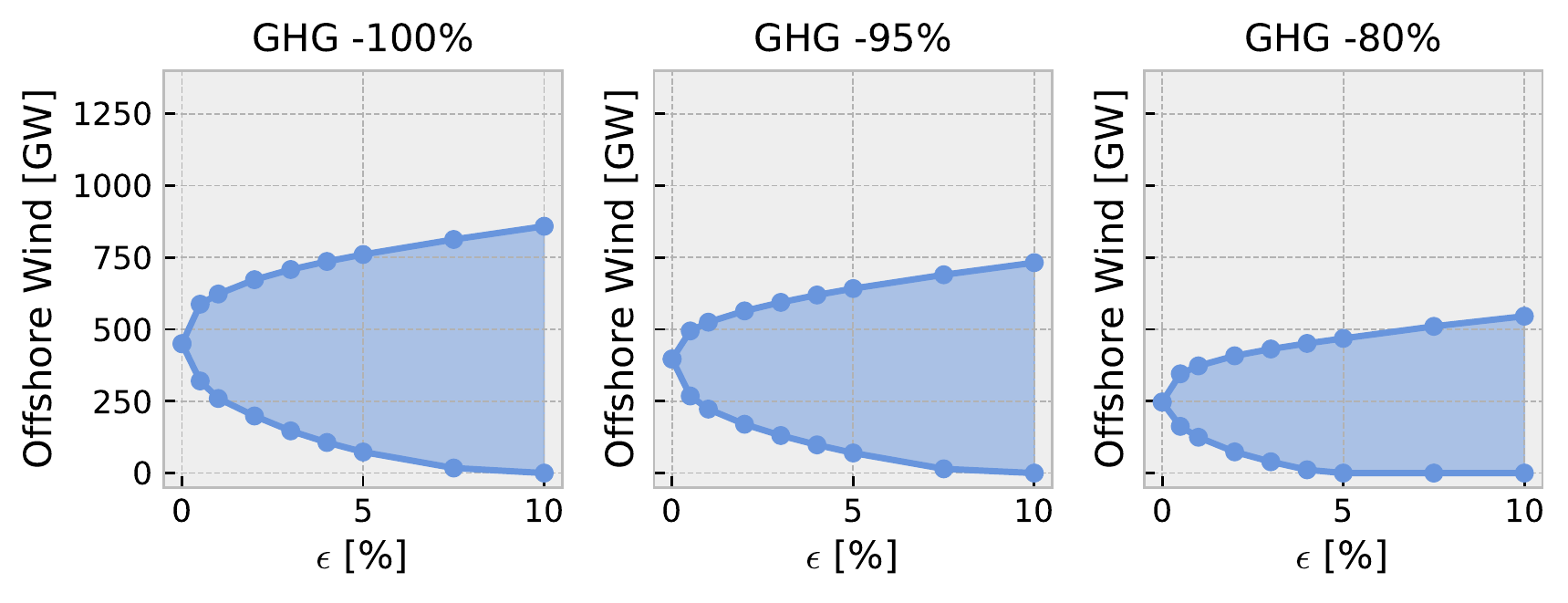}
  \includegraphics[width=\columnwidth]{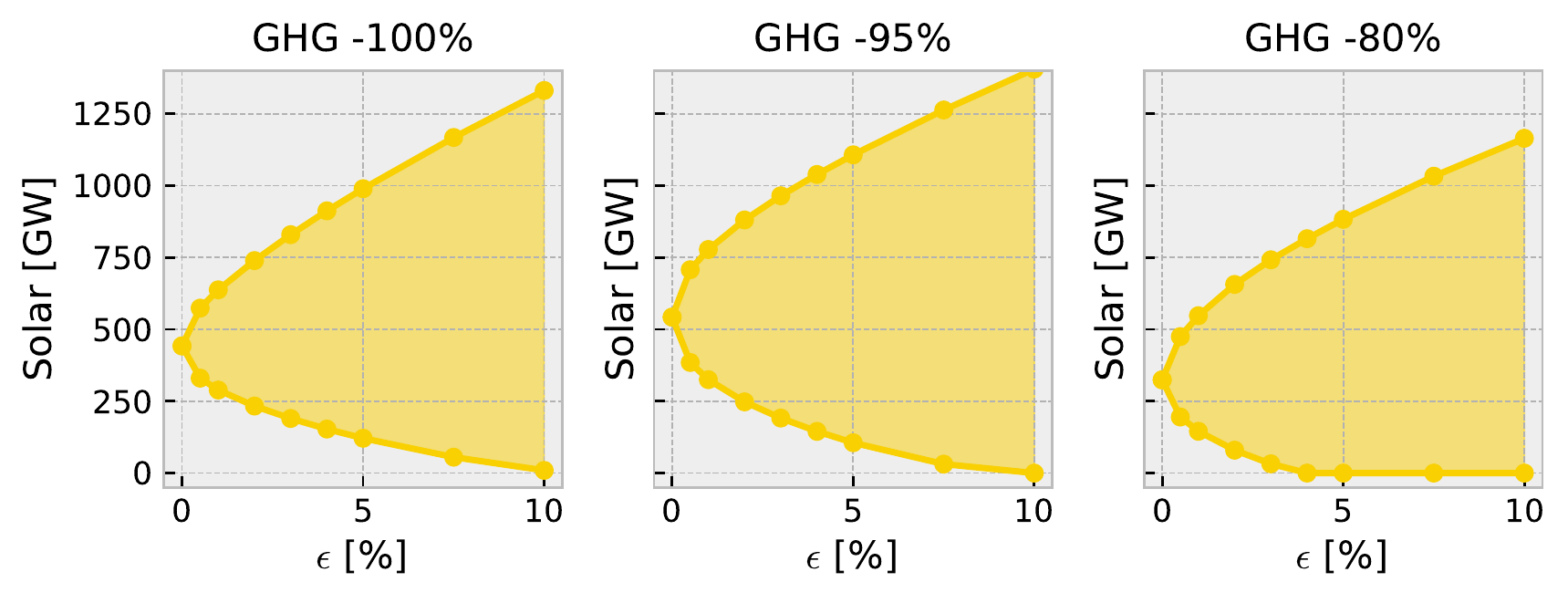}
  \includegraphics[width=\columnwidth]{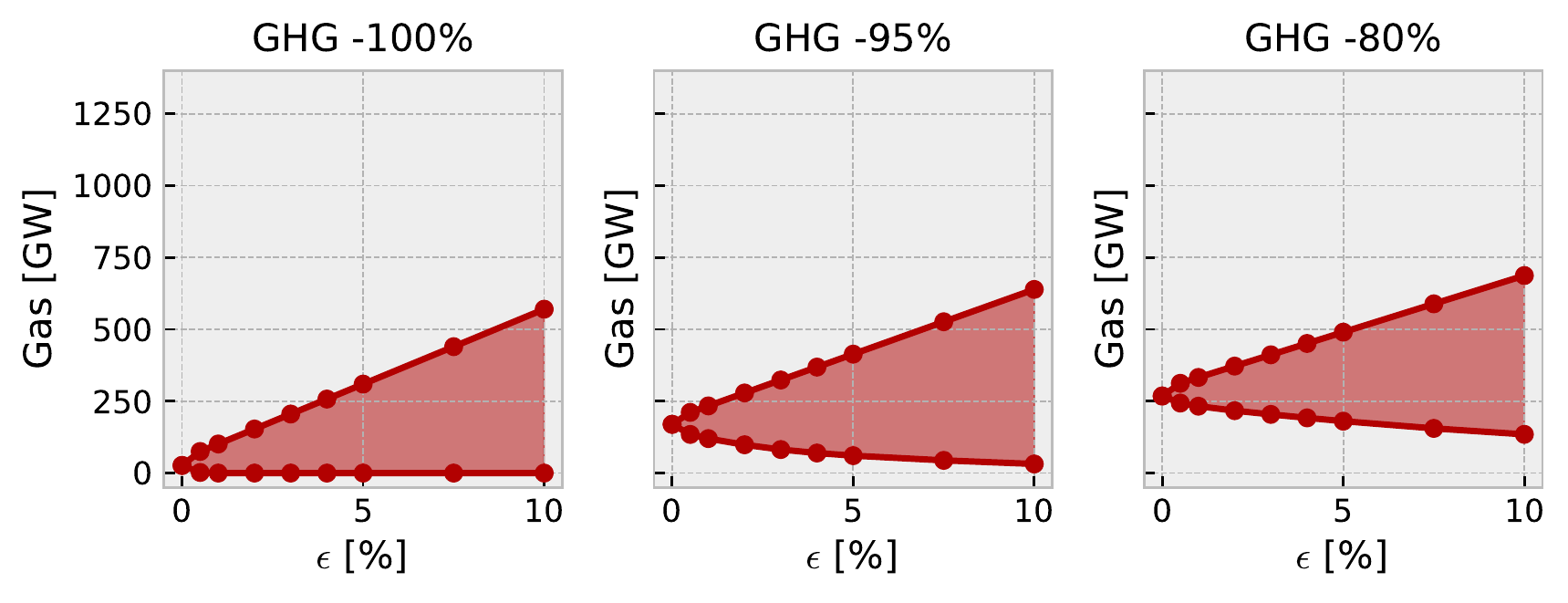}
  \vspace{-0.6cm}
  \caption{Solution space of renewable generation infrastructure by technology for different levels of slack $\epsilon$ and emission reduction targets.}
  \label{fig:space-generation}
  \vspace{-0.3cm}
\end{figure}

In this section we extremise different groups of investments in generation, storage and transmission infrastructure.
As an example, Fig. \ref{fig:maps-tmin}-(ii) depicts a system that seeks to deviate from the
optimum by minimizing the volume of transmission network expansion up to a total cost increase
of 10\%, for instance as a concession to better social acceptance.
With the results particularly the NordSued link connecting Northern and Southern Germany
manifests as a no-regret investment decision up to a capacity of 15 GW in the context of full decarbonization.
It is one of the few persistent expansion routes.
All other transmission expansion corridors are (to a significant extent) not compulsory.
Missing transmission capacities can be compensated by adding storage capacity and more regionally dispersed power generation.
Nevertheless, some transmission network reinforcement is indispensable to remain within the given cost bounds.
These results are also broadly aligned with findings in
\cite{schlachtberger_benefits_2017, Hoersch2017d}.

Beyond this example, the MGA results offer insights about the structure of the near-optimal space.
The intent is to portray a set of technology-specific rules that must
be satisfied to keep costs within pre-defined ranges $\epsilon$.
Note, that the discontinuity created by $\epsilon$ restricts the accuracy of the solution space.

Fig. \ref{fig:space-generation} reveals that
wind generation, either onshore or offshore, is essential to set up a cost-efficient European
power system for all three evaluated emission reduction levels.
Whilst already a small cost increase of 0.5\% yields investment flexibilities in
the range of $\pm$100 GW, even a 10\% more costly solution would still require more
than 500 GW of wind generation capacity for a fully renewable system:
two-thirds of the optimal system layout.
However, even for a zero-emission system a cost increase of just 4\% enables
abstaining from onshore wind power, and a 7.5\% more expensive alternative
can function without offshore wind farms.

The investment flexibility develops non-linearly with increasing slack levels $\epsilon$.
Even a minor deviation from the cost optimum by 0.5\% creates room for maneuver
in the range of $\pm$200 GW for onshore and $\pm$150 GW offshore wind installations,
which indicates a weak tradeoff between onshore and offshore wind capacities
very close to the optimum. Nonetheless, dispensing with both is not viable.
Furthermore, 10\% of total system costs must be spent to rule out solar panels,
but already a slack of 1\% allows to reduce the solar capacity by a third.

Price et al. observed in their model that investment flexibility in generation
infrastructure decreased as more tight caps on carbon-dioxide emissions were
imposed \cite{Price2017}. While it is true that more ambitions climate protection
plans incur more must-haves (i.e. minimum requirements of capacity), for the
case-study at hand the viable ranges of marginally inferior solutions
increase as more total capacity is built.

\begin{figure}
  \centering
  \includegraphics[width=\columnwidth]{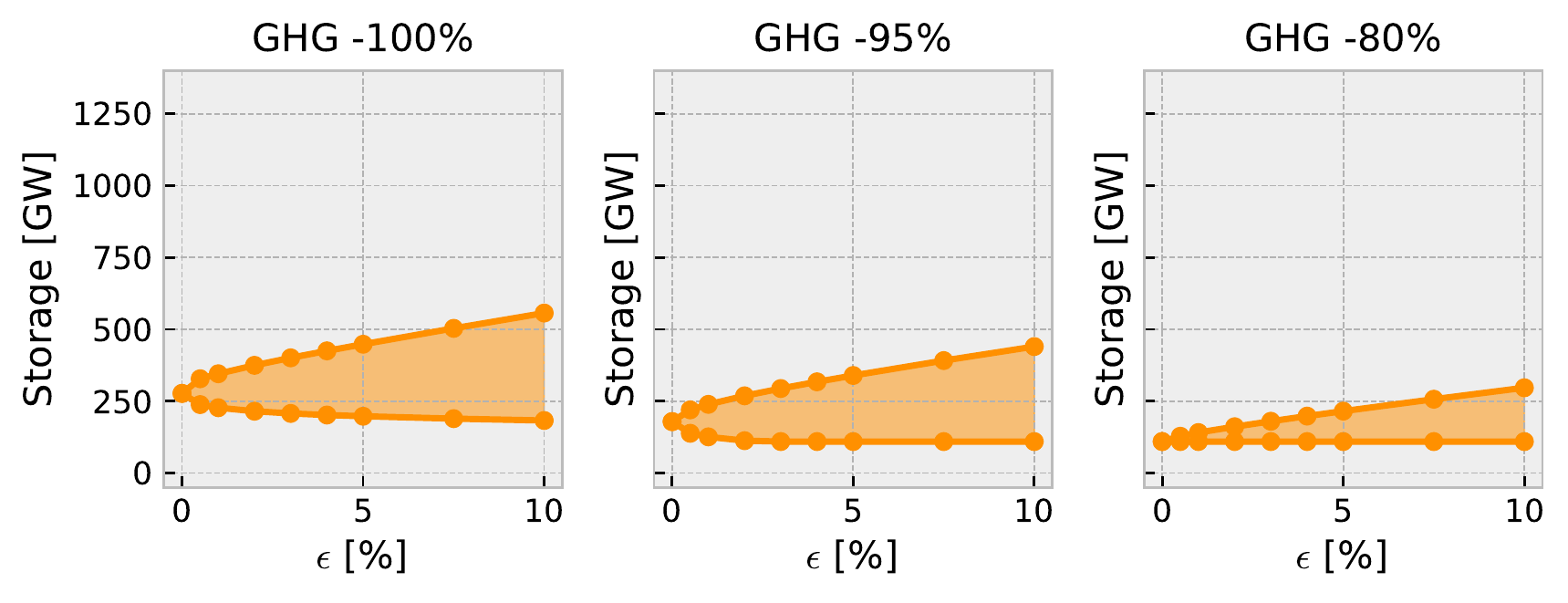}
  \includegraphics[width=\columnwidth]{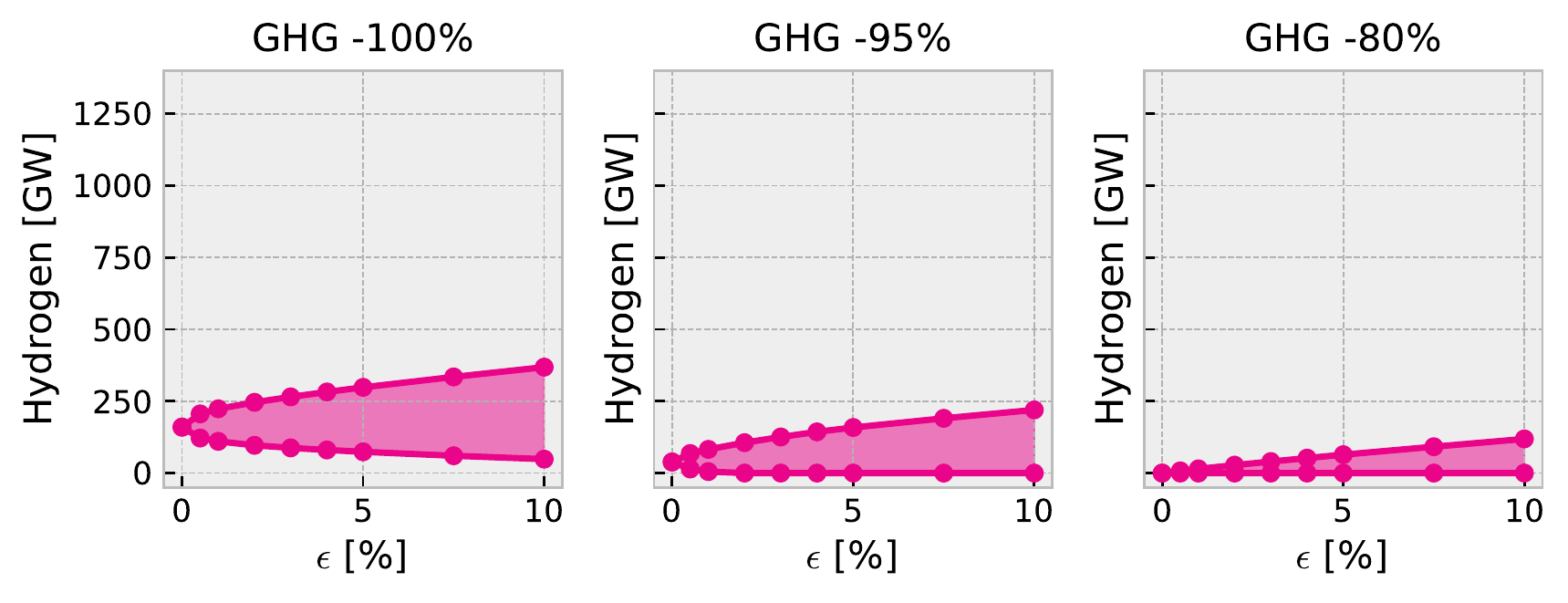}
  \includegraphics[width=\columnwidth]{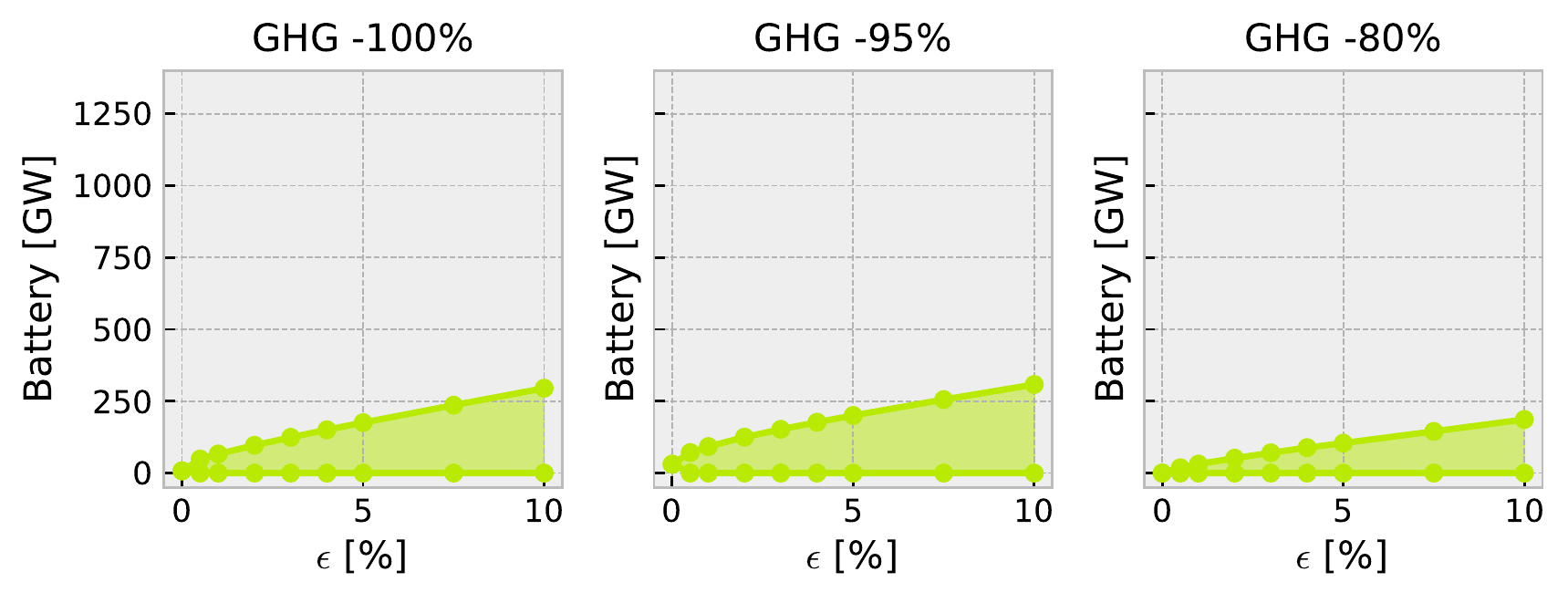}
  \includegraphics[width=\columnwidth]{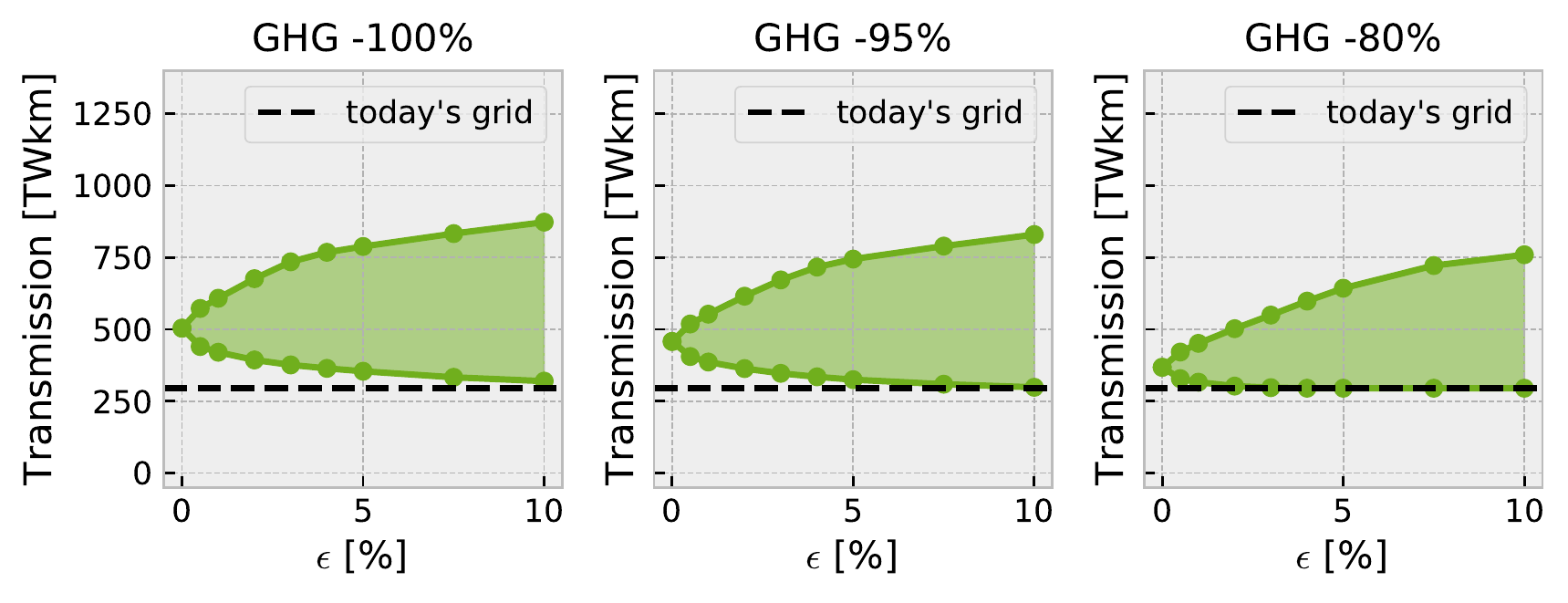}
  \vspace{-0.6cm}
  \caption{Solution space of storage and transmission infrastructure by technology for different levels of slack $\epsilon$ and emission reduction targets.}
  \label{fig:space-storage}
  \vspace{-0.3cm}
\end{figure}

As Fig. \ref{fig:space-storage} exhibits, even for a complete decarbonization
of the European power system building battery storage is not essential, although
they are deployed in response to e.g. minimizing network reinforcement.
Conversely, once weather-independent dispatch flexibility from natural-gas-fired power
plants is unavailable, it becomes imperative to counter-balance with hydrogen storage.
The cutback of hydrogen infrastructure under these circumstances goes along with
building excess generation capacities and multiplied amounts of curtailment.

The reinforcement of the transmission network becomes more pivotal the
more the power system is based on renewables.
Aiming for an emission reduction by 80\% a 2\% more expensive variant can get by without any grid reinforcement.
Reducing emissions by 100\% still requires some additional power transmission capacity at a 10\% cost deviation.
However, within this range, the transmission volume can deviate from almost
double of today's network capacities to merely a marginal reinforcement
(cf. Fig. \ref{fig:maps-tmin}).

MGA iterations were also applied from a country-wise perspective.
Remarkably, any one country could completely forego any one generation or storage
technology and remain within 5\% of the cost optimum when targeting a 95\% reduction
in greenhouse gas emissions. In this case, neighboring countries must offset
the absence of this technology.

\begin{figure}
  \centering
  \includegraphics[width=0.87\columnwidth]{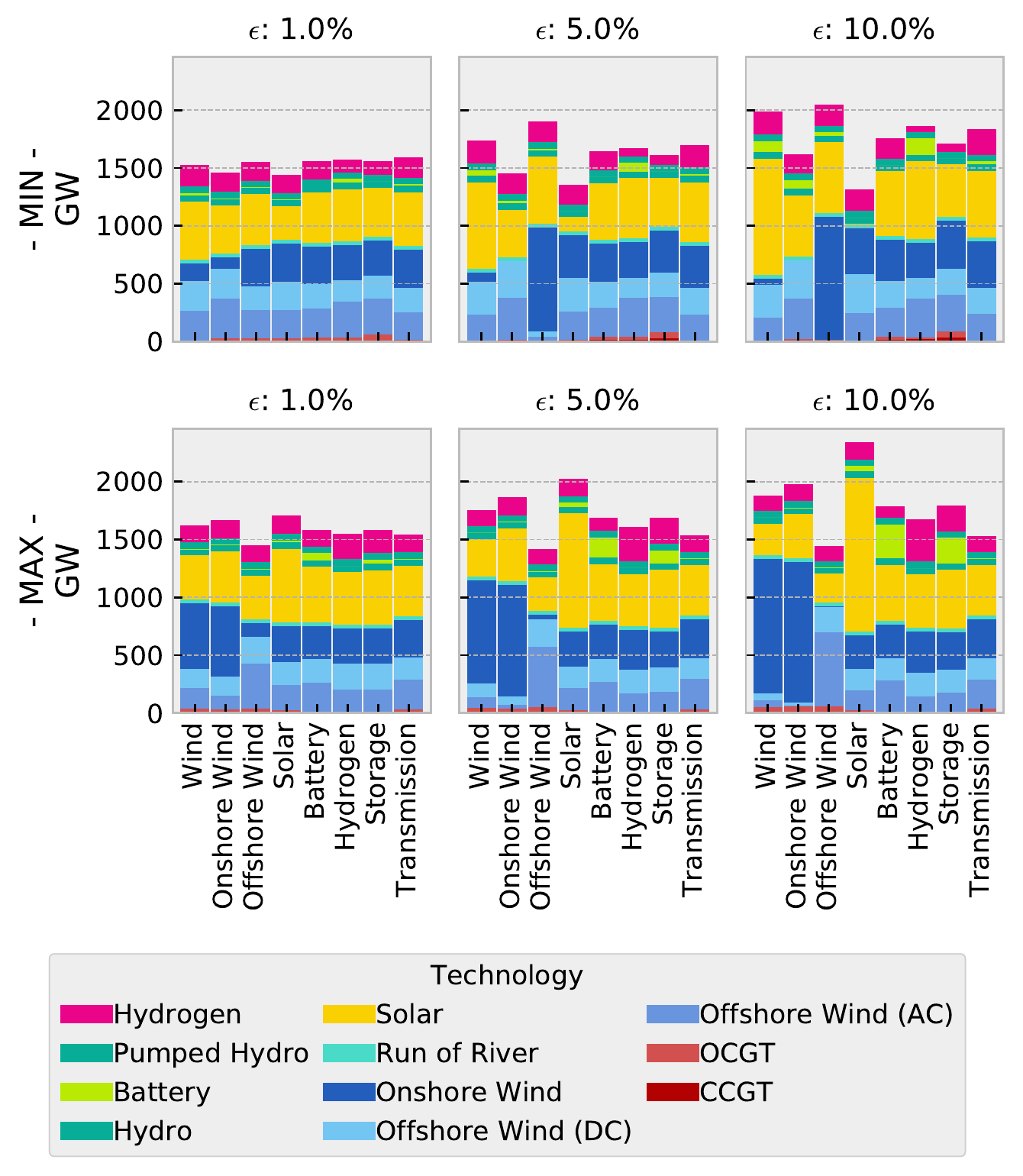}
  \caption{Composition of generation and storage capacities for various near-optimal solutions
    with 100\% renewables.
    Each subplot corresponds to a slack level $\epsilon$ and an optimization sense. The labels of the bar charts indicate which group of investment variables is included in the objective.}
  \label{fig:capmix}
  \vspace{-0.3cm}
\end{figure}

\begin{figure}
  \centering
  \includegraphics[width=0.67\columnwidth]{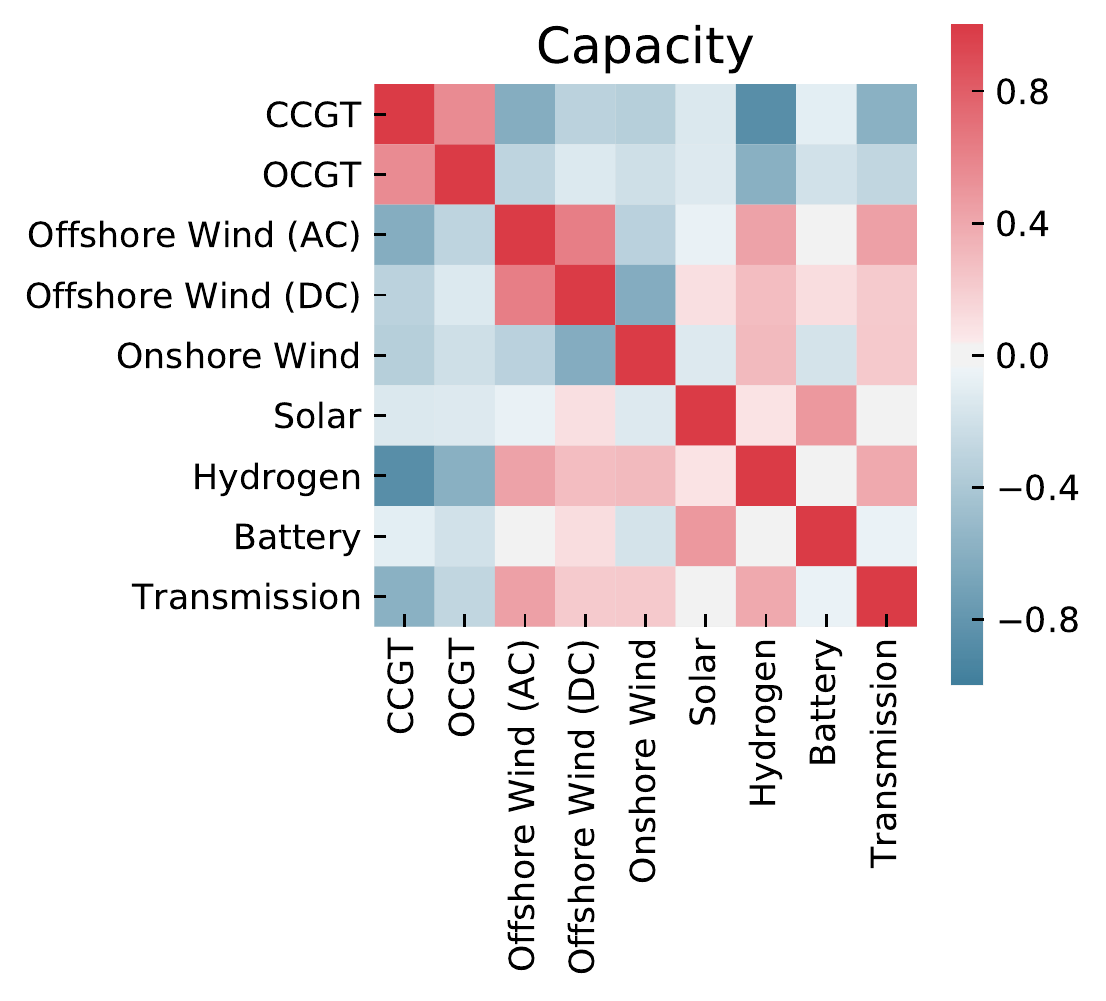}
  \caption{Correlations of capacities across all near-optimal solutions.}
  \label{fig:correlations}
  \vspace{-0.4cm}
\end{figure}

\subsection{Correlations}

So far, the study of the near-optimal feasible space did not capture the interdependence
between different system components apart from the envelopes the analysis provided for each technology.
Shifting to the extremes of one technology will diminish the investment flexibility of other carriers.
Fig. \ref{fig:capmix} demonstrates how diversity in capacity mixes rises if more leeway is given in terms of system costs.
The striking variety in capacity totals is largely attributable
to the lower capacity factors of solar compared to wind energy.

Hennen et al. suggested to present the intertwining of technologies through
Pearson's correlation coefficient across all near-optimal solutions \cite{Hennen2017}.
Fig. \ref{fig:correlations} confirms many of the previously noted connections.
Hydrogen storage substitutes natural gas turbines and is positively correlated with
onshore and offshore wind capacity, while battery deployment rather matches with solar installations.
Likewise, transmission expansion occurs in unison with onshore and offshore wind deployment.
Thereby, hydrogen storage and transmission become complements for high renewable energy scenarios.
With caution should be noted that CCGT and OCGT as well as
AC- and DC-connected offshore wind installations have high correlations
because they are grouped in the MGA iterations.

\subsection{Distributional equity}

Surveys suggest that an equal distribution of power supply is preferred among the population
and may increase their willingness to participate in the energy system transformation process \cite{drechsler_efficient_2017}.
Sasse et al. and Drechsler et al. investigate the trade-offs between least-cost and
regionally equitable solutions in Switzerland and Germany by using concepts of the
Lorenz curves and the Gini coefficients as equity measures \cite{sasse_distributional_2019, drechsler_efficient_2017}.

\begin{figure}
  \centering
  \includegraphics[width=0.6\columnwidth]{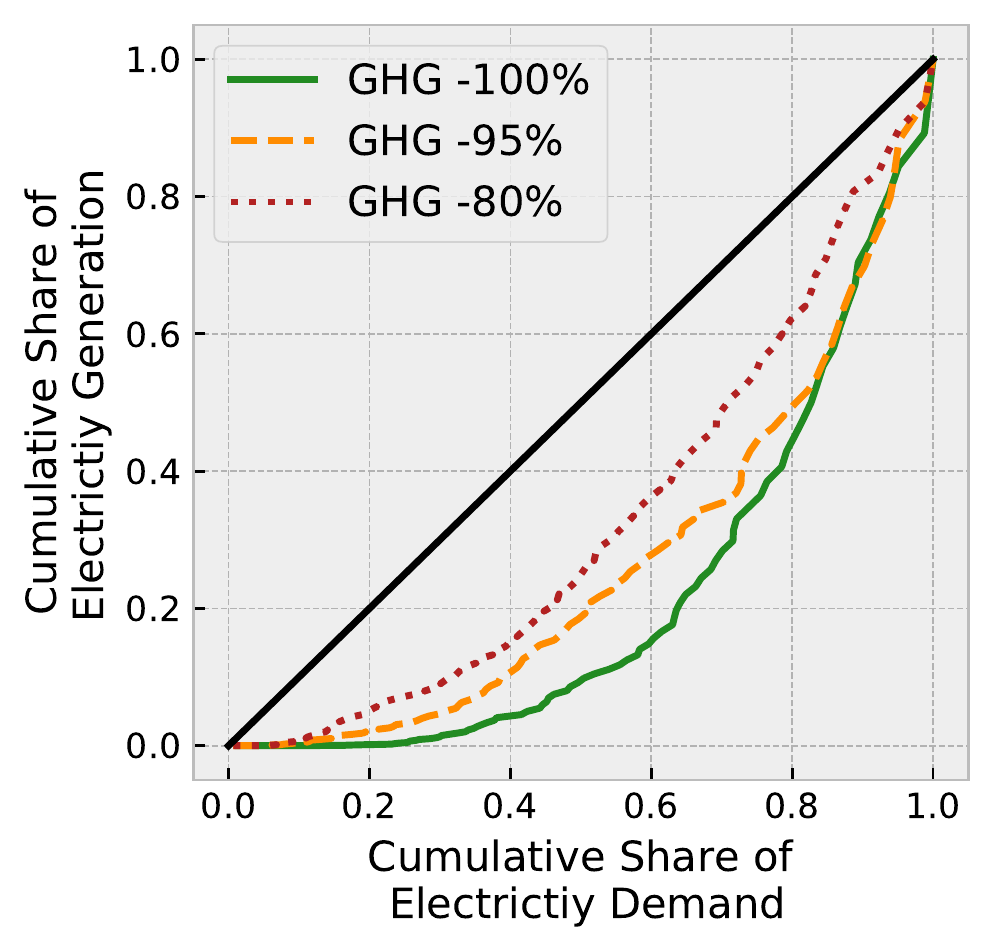}
  \caption{Lorenz curves for the optimal solutions at different emission reduction levels
    relating the cumulative share of electricity generation to the cumulative share of demand
    in the 100 regions of the European power system model.}
  \label{fig:lorentz}
  \vspace{-0.3cm}
\end{figure}

In the context of power, the Lorenz curve can relate the cumulative share of electricity generation
of regions to their cumulative share of electricity demand as shown in Fig. \ref{fig:lorentz}.
For more ambitious emission reduction targets, less equitable solutions are favored from a
cost-perspective, however, they may not be in line with the public attitude.

The Gini coefficient is the corresponding scalar measure of uniformity and is
determined by multiplying the area between the Lorenz curve and the identity line by two.
A Gini coefficient of 1 represents the most unequal distribution,
while 0 corresponds to the situation where every region produces, on average, as much electricity as they consume.

\begin{figure}
  \centering
  \includegraphics[width=\columnwidth, trim=0.25cm 0cm 0.3cm 0cm, clip]{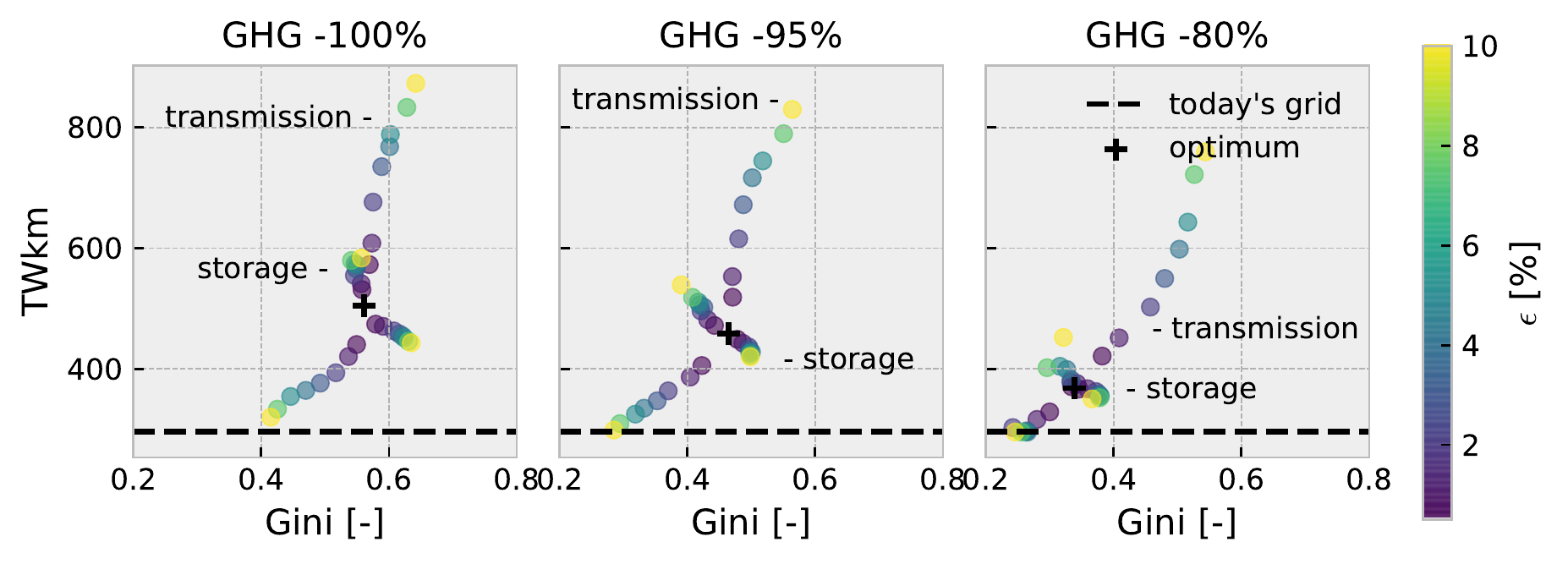}
  \caption{Development equitable power generation measured by the Gini coefficient in relation to the capacity of the power transmission network measured in TWkm. The four plotted search directions are minimization and maximization of each, total transmission volume and storage capacity.}
  \label{fig:search-directions}
  \vspace{-0.4cm}
\end{figure}

Preceding studies have, in general, noted that the focus on wind power tends
to be detrimental to a regionally balanced distribution of electricity generation
\cite{drechsler_efficient_2017}, whereas photovoltaic power supply is a key factor
for an even distribution of the power supply \cite{sasse_distributional_2019}.
This observation is consistent with the results of the European power system at hand.
But as we consider the transmission grid and energy storage options,
we can further extend on these findings.
In Fig. \ref{fig:search-directions} we track the equity of solutions in relation to
the transmission network volume as we approach the boundaries of either technology category.
Substantially more regionally equitable solutions are attainable for a limited
cost-increase when diverting attention away from transmission network expansion.
Utilizing less energy storage rather discourages equitable generation patterns;
in the opposite direction, this is not the case for a zero-emission system.
Note that the equity measures are only an observed side-effect and not the
objective of a particular search direction of the MGA method.
There is no guarantee that no more equitable solutions exist within the near-optimal feasible space.

\section{Critical Appraisal}
\label{sec:appraisal}

This paper covers the electricity sector only. Brown et al.\ suggested that
with an increasing coupling of energy sectors the benefit of the transmission system decreases \cite{Brown2018b}.
It is not far-fetched, that the near-optimal feasible space,
in general, might look very different with a tightened sectoral integration.

Within the computational constraints, it is moreover desirable to further
enhance the spatial and temporal resolution to better reflect curtailment
caused by transmission bottlenecks and factor in extreme weather events \cite{Hoersch2017d}.

This work further neglects parametric uncertainty. Coupling this paper's variant of MGA
with a parameter sweeping method (such as Monte Carlo simulation \cite{Li2017})
would allow to derive a more sophisticated
version of Fig. \ref{fig:space-generation} and \ref{fig:space-storage} with fuzzy boundaries
that represent the probability with which the respective capacities of system components
are contained within the near-optimal feasible space.

\section{Summary and Conclusions}
\label{sec:conclusion}

This work sheds light on the flatness of the near-optimal feasible decision space
of a power system model with European scope for ambitious climate protection targets.

An understanding of alternatives beyond the least-cost solutions
is indispensable to
develop robust, credible and comprehensible policy guidelines.
Therefore, we derived a set of technology-specific boundary conditions
that must be satisfied to keep costs within pre-defined ranges using the
modeling-to-generate-alternatives methodology.
These rules permit well-informed discussions around social constraints
to the exploitation of renewable resources or the extent to which the power
transmission network can be reinforced in discussions.

Indeed, we observed high variance in the deployment of individual system components,
even for a fully renewable system. Already a minor cost deviation of 0.5\% offers
a multitude of technologically diverse alternatives.
It is possible to dispense with onshore wind for a cost increase of 4\%,
and to forego solar for 10\%.
Nevertheless, either offshore or onshore wind energy plus at least some
hydrogen storage and grid reinforcement are essential to keep costs within 10\% of the optimum.

\balance
\section*{Supplementary Material}

For supplementary material and reading the reader is referred to the
source code repository on Github (\url{https://github.com/pypsa/pypsa-eur-mga})
and the documentations of PyPSA (\url{https://pypsa.readthedocs.io}, \cite{brown_pypsa_2018})
and PyPSA-Eur (\url{https://pypsa-eur.readthedocs.io}, \cite{Horsch2018}).




\bibliographystyle{IEEEtran}
\bibliography{library.bib}
%


\end{document}